\documentclass{mem}
\usepackage{natbib}
\usepackage{txfonts}
\usepackage{balance}
\usepackage{graphicx}
\usepackage[latin1]{inputenc}
\usepackage[a4paper,breaklinks,dvipdfm]{hyperref}
\idline{75}{1}
\begin{document}
\def\teff{$T\rm_{eff }$}
\def\kms{$\mathrm {km s}^{-1}$}

\title{
Observing strategy of the THESEUS mission}

\subtitle{}

\author{
F. \, Frontera\inst{1,2} \and L. \, Amati\inst{1} \and P. \, O'Brien\inst{3} \and D. \, G{\"o}tz\inst{4} \and E. \, Bozzo\inst{5} \and C. \, Tenzer\inst{6} \and R. \, Campana\inst{1} \and F. \, Fuschino\inst{1} \and C. \, Labanti\inst{1} \and M. \, Orlandini\inst{1} \and P. \, Attinà\inst{7} \and C. \, Contini\inst{8} \and B. \, Morelli\inst{8} \\
\hspace{0.5cm}\\
on behalf of the THESEUS Consortium
}
\institute{
Istituto Nazionale di Astrofisica --
OAS, Via Gobetti 101,
I-40139 Bologna, Italy
\and
Ferrara University, Department of Physics and Earth Sciences,
Via Saragat 1, 44122 Ferrara, Italy, \email{frontera@fe.infn.it}
\and 
University of Leicester, Department of Physics and Astronomy, Leicester, LE17RH, UK
\and
IRFU/Département d'Astrophysique, CEA, Université Paris-Saclay, F91191, Gif-sur-Yvette, France
\and
University of Geneva, Department of Astronomy, ch. d'Ecologia 16, CH-1290 Versoix, Switzerland
\and
Institut f{\"u}r Astronomie und Astropysik, Universit{\"a}t T{\"u}bingen, Germany
\and
Istituto Nazionale di Astrofisica, OATo, Torino, Italy
\and
OHB Italia, Roma
}
\authorrunning{Frontera}
\titlerunning{Observing strategy}
\abstract{
We will discuss the observing strategy of the {\em Transient High Energy Sky and Early Universe Surveyor} (THESEUS) mission proposed to ESA as a response to the M5 call for proposals. The description of THESEUS and its science goals can be found in the white paper by \citet{Amati17}. 
}
\maketitle{}

\section{Introduction}

The description of THESEUS and its science goals can be found in the white paper by \citet{Amati17}. The main requirements of the observing strategy can be summarized as follows:
\begin{itemize}
\item
Discover at low energies (0.3--6~keV) an unprecedented high number of GRBs (especially at high redshifts), as well as a wide range of astropysical transient sources or new events in the local and remote Universe.
\item
Localize the detected new transient events with 0.5--1 arcmin accuracy;
\item
Identify the nature of the detected transients via on-board catalogs and/or high energy ($>2$~keV)imaging and/or spectroscopy.
\item
Promptly follow-up the secure GRBs or new transient phenomena in the Near Infrared (NIR) band and determine their position with arcsecond accuracy, ad perform NIR spectroscopy and/or photometry;
\item
Promptly comunicate new triggers to the ground in order to enable ground/space observatories to follow-up the new transients;
\item
Ensure the needed pointing stability of the satellite for at least 1~ks in order to detect the longest events while the detected events are observed in the NIR band.
\end{itemize}

In order to achieve these goals the following payload instrumentation is developed and adopted (see Fig.~\ref{f:payload}):

\begin{figure}[htbp]
\centering
\includegraphics[width=0.50\textwidth]{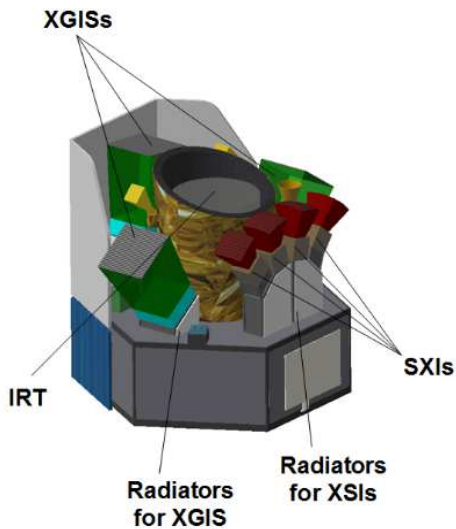}
\caption{\footnotesize THESEUS payload.}
\label{f:payload}
\end{figure}

\begin{figure}[htbp]
\centering
\includegraphics[width=0.50\textwidth]{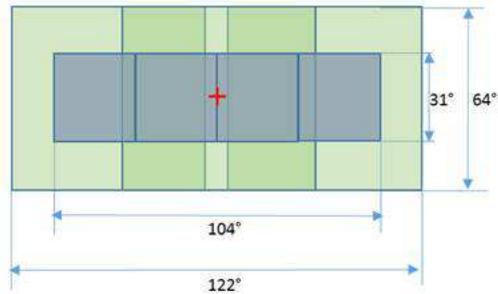}
\caption{\footnotesize
FOV of the SXI ({\em inner field}) and XGIS ({\em outer field}).}
\label{f:FOV}
\end{figure}

\begin{itemize}
\item
A Soft X--ray Imager (SXI) based on 4 Lobster--Eye telescopes with 0.3--6~keV energy band, a 0.5--1~arcmin location accuracy and a Field of View (FOV) of about 1~sr;
\item
An X--/Gamma--ray Imaging Spectrometer (XGIS) based on 3 coded-mask telescopes with 2~keV--20~MeV energy band, similar sensitivity to the SXI in the common energy band (2--6~keV), 5 arcmin positioning accuracy up to 50 keV, FOV covering the SXI FOV (see Fig.~\ref{f:FOV}).
\item
An InfraRed Telescope (IRT) of 70~cm aperture, a Cassegrain geometry, 0.7--1.8~$\mu$m wavelength band, 10'$\times$10' FOV, sensitivity H$=$20.6~AB mag (300~s).
\item
IRT observation condition of the GRB localized with SXI: when the satellite stabilization is within 1~arcsec.
\item
At a sensitivity limit H$=$18.5 (AB), 5 minute integration time is sufficient for a Low Resolution spectrum with IRT.
\end{itemize}

\section{Observing strategy}

\subsection{How to localize new transient events with SXI} 

\begin{figure}[htbp]
\centering
\includegraphics[width=0.50\textwidth]{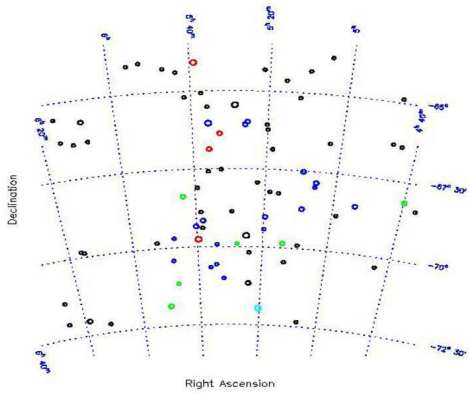}
\includegraphics[width=0.50\textwidth]{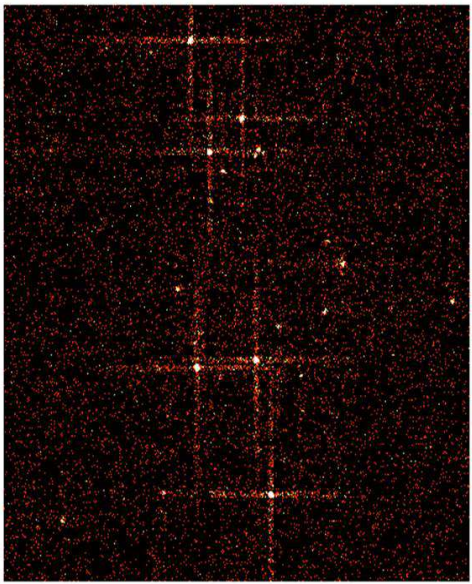}
\caption{\footnotesize
{\em Top}: X--ray map of the Large Magellanic Cloud (LMC). {\em Bottom}: LMC image  as expected from a Lobster-Eye Telescope.}
\label{f:LMClocation}
\end{figure}

The location accuracy of the SXI is illustrated in 
Fig.~\ref{f:LMClocation}, in which it is simulated the capability of a Lobster-Eye telescope in the location of X--ray sources in the Large Magellanic Cloud (LMC).

The procedure we intend to adopt for the selection of  new transient events with SXI is the following:
\begin{itemize}
\item
Apply an algorithm to the SXI data in order to get an event list with position (in CCD pixels) and pulse height, for different integration times (2, 20, 200, 2000~s);
\item
Convert pixel positions into a local coordinate frame;
\item
Subtract background pattern from the histograms;
\item
Scan the histograms for significant peaks and extract candidate positions;
\item
Calculate accurate positions in the local frame;
\item
Transform the positions in global sky coordinates;
\item
Check SXI positions against onboard catalogues of known sources;
\item
Communicate unknown transients detected with SXI to the On Board Data Handling (OBDH), as shown in Fig.~\ref{f:OBDH}.
\end{itemize}

\subsection{Identification of the transient events to be followed-up by the IRT}

As the SXI telescopes can be triggered by many classes of transient phenomena  (e.g., flare stars, X-ray bursts, GRBs,  etc.), the XGIS provides an efficient means to help the identification of the high energy transient phenomena (GRBs, Soft Gamma-ray Repeaters, Tidal Disruption Events, X-ray counterparts of Fast Radio Bursts, etc.). Thus, in parallel to the selection done with SXI Data Handling Unit (see Fig.~\ref{f:OBDH}), the following trigger conditions are searched:

\subsubsection{XGIS trigger condition based on images}

\begin{itemize}
\item
For each XGIS unit, the current 2--30 (or 50)~keV images integrated over 3 time durations (typically 1, 10, 100~s) is continuously compared with the corresponding reference images obtained by averaging the previous (typically 30) images;
\item
A trigger condition is satisfied if a spot in the images emerges at a significance level of at least $n\sigma$ (typically $n=5$).
\end{itemize}

\subsubsection{XGIS trigger condition based on data rate}

\begin{itemize}
\item
Continuous monitoring of the data rate in the 2--30~keV and 30--200~keV energy bands for each of the 4 modules that make each of the 3 XGIS units;
\item
The trigger condition is satisfied when, in one or both energy ranges, at least 3 detection modules see a simultaneous excess at a significance level of at least $n\sigma$ (typically $n=5$) on at least one of 4 different time scales (e.g., 10~ms, 100~ms, 1~s, 10~s).
\end{itemize}

\begin{figure}[htbp]
\centering
\includegraphics[width=0.50\textwidth]{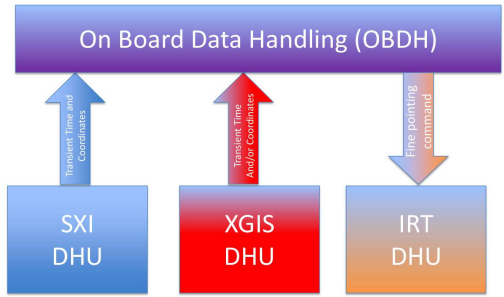}
\caption{\footnotesize
On board logic for commanding an IRT follow-up observation.}
\label{f:OBDH}
\end{figure}

Once the time of the XGIS event direction (from images) and/or that of the data excess are sent to the OBDH (see Fig.~\ref{f:OBDH}), the OBDH logic selects the events to be followed--up by the IRT. These can be events that have been recognized to be unknown transients with SXI (see above) or events imaged with the XGIS that satisfy all the XGIS trigger conditions.

\section{How to promptly down-link the trigger results}
\label{s:downlink}

The baseline for promptly transmitting (less than 4 min) the trigger data to ground exploits the VHF network adopted for the SVOM mission \citep{Wei16}. In any case, during the Phase A study, other possibilities will be investigated:
\begin{itemize}
\item
Exploitation of the {\em Iridium} satellite constellation now in advanced development {\em Iridium Next}, that would allow to transmit the trigger data in 1 min.
\item
Exploitation of the {\em European Tracking and Data Relay Satellite System} (TDRSS) now under development, that would allow to transmit the trigger data in less than 1 min.
\item
American TDRSS ($<1$~min) in the case of an international participation by NASA.
\end{itemize}

\section{Predicted detection rate of GRBs and their properties as determined with THESEUS}
We expect the following:
\begin{itemize}
\item
As far as the detection rate, we expect a number of GRBs between 300 and 700 events/year.
\item
For most of the distant GRBs, we expect to be capable to determine their photometric redshift with the IRT, while for the strongest ones we expect to determine their spectroscopic redshift.
\item
Those GRBs, for which the photometric redshift will be 
determined on board, are the best candidates for study with large ground/space facilities, like E-ELT, ATHENA, etc.
\item
We expect that THESEUS will be capable to provide more spectroscopic redshifts in one year than {\em Swift} in a decade.
\end{itemize}  

\section{How to manage external triggers or TOO requests}

The external triggers, that will be up-loaded via Malindi ground station or via the same system adopted for the downlink (see Section~\ref{s:downlink}), will arrive to the OBDH and will be managed using the same logic adopted for the internal triggers (see Fig.~\ref{f:OBDH}).
 
\section{Constraints on the THESEUS solar attitude angle and sky exposure map in one year}

\begin{figure}[htbp]
\centering
\includegraphics[width=0.50\textwidth]{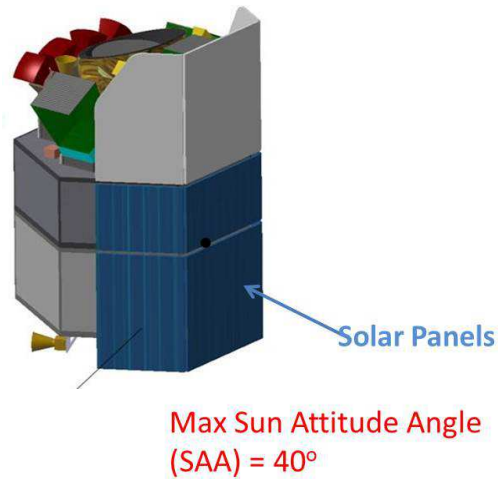}
\includegraphics[width=0.50\textwidth]{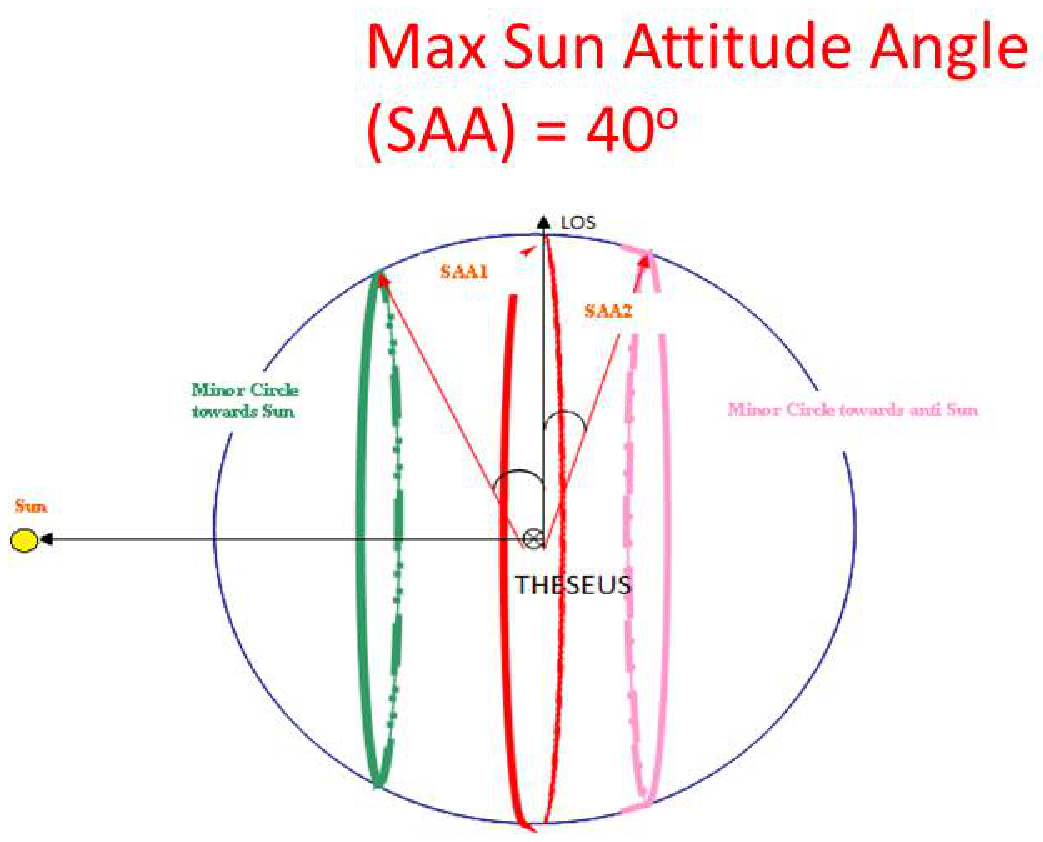}
\caption{\footnotesize
{\em Top}: THESEUS solar panels folded at the satellite launch. {\em Bottom}: Range of Sun Attitude Angles (SAA) compatible with the required power.}
\label{f:SAA-Theseus}
\end{figure}

When the satellite is in Sunlight, the solar panels (see left panel of Fig.~\ref{f:SAA-Theseus}) provide the electrical power to all THESEUS subsystems, keeping charged the satellite batteries. In order to guarantee  the needed power, the maximum offset angle between the solar panel axis and the Sun direction has to be at most $\pm 40$~degrees (see right panel of Fig.~\ref{f:SAA-Theseus}).

When the satellite is in the Earth shadow, other possible attitudes are in principle possible, including the Anti-Sun direction. The attitude strategy will be deeply investigated during the Phase A study.

\begin{figure}[htbp]
\centering
\includegraphics[width=0.50\textwidth]{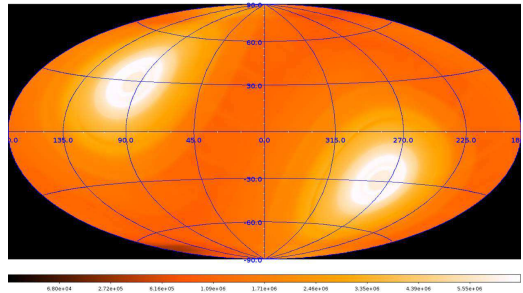}
\caption{\footnotesize
Exposure map of the sky in on year, assuming the attitude constraints adopted in Sunlight (see Fig.~\ref{f:SAA-Theseus})}.
\label{f:skymap}
\end{figure}

Assuming the attitude constraints adopted in Sunlight, the exposure map of the sky in 1 year is shown in Fig.~\ref{f:skymap}.

\section{Ground Segment}

As for other ESA missions with similar profiles, it is expected that the THESEUS ground segment will comprise a Mission Operation Center (MOC) and a Science Operation
Center (SOC), both led by ESA. 

The MOC is usually the sole interface with the satellite, managing all spacecraft operations, uploading the telecommands, controlling the downlink of the telemetry and
collecting it from the mission ground stations, as well as performing all necessary maintenance and emergency operations required to guarantee the safety and nominal functioning of the spacecraft.

The SOC is usually in charge of the communications with the world-wide community, issuing the call for proposals, organizing the peer reviewed selections and compiling the yearly-long observational
timeline. Being THESEUS a mission focused on GRBs and transient sources, the core of the observing program is expected to be dominated by the follow-up of these events (typical observational campaigns might comprise tens of few ks long pointings). Thus an important task will rely on the ESA appointed project scientist, who will have the final decision concerning the implementation (or not) of the
target of opportunity observations (ToOs) that will be requested by the community. The largest fraction of the ToOs are expected to derive from the
SXI and XGIS triggers, but, as mentioned above,  a large fraction of external triggers is also expected from the other facilities operating also at different wavelengths.

For the latter, as well as for the ToOs, detailed planning
constraints (e.g., sky visibility, and mission resources such as power and telemetry) will need to be identified by the SOC on a daily
time scale in order to shape the short term observational schedule. A similar task will have to be covered by the MOC for what concerns the payload operations exclusion windows,
on-board resources envelopes for payload operations, and final detailed checks against mission, environmental and resource constraints.

The basic Mission Planning approach for all the routine science operations phases will be built on the experience
of previous missions, such as XMM-Newton, Herschel and INTEGRAL. The rapid rescheduling imposed by the GRB nature of the mission will take
advantage of the heritage gained through the {\em Swift} mission. A THESEUS science data center (SDC) will be provided by the consortium to support the MOC and the SOC
in the routinely processing, inspection, and validation of the data. The SDC is also expected to provide the required software for scientific analysis, as well as the
pipelines needed for the daily data processing. 

There may be different instrument operation centers (IOCs): for the XGIS, the SXI, and the IRT. The IOCs will provide full support to ESA
for the payload integration and operations, and will generally host and maintain the instrument expertize all along the mission lifetime (including the post operations).
They will reasonably also provide updated calibrations for each instrument whenever required.  

It is expected that a centralized archive will be hosted at SOC,
functioning as the primary repository for all science data products, calibrations, documentation, and software. This will guarantee the long term exploitation of the
mission data and guarantee a proper mission heritage. During the assessment phase, all available resources for the ground segment will be analyzed
to optimize the ground segment design and maximize the scientific return of THESEUS.

\section{Conclusions}
The main science goals of the mission (the most demanding) can be achieved using the described strategy. 

Other technical constraints, like instrument temperature stability, attitude stability and accuracy, OBDH logic reliability, reliability of the re-pointing automatic system for a fast NIR observation, are crucial to enable the discussed observing strategy. 
They have been discussed in the white paper by \citet{Amati17}.

\begin{acknowledgements}
We are grateful to all members of the Theseus Consortium for stimulating the definition of the  observing strategy constraints.
\end{acknowledgements}

\bibliographystyle{aa}

\begin{thebibliography}
\expandafter\ifx\csname natexlab\endcsname\relax\def\natexlab#1{#1}\fi
\bibitem[Amati et al.(2017)]{Amati17} 
Amati, L., O'Brien, P., Goetz, D., et al.\ 2017, arXiv:1710.04638 
\bibitem[Wei et al.(2016)]{Wei16} Wei, J., Cordier, B., Antier, S., et al.\ 2016, arXiv:1610.06892 
\end{thebibliography}

\end{document}